\documentclass[twocolumn, prb, showpacs]{revtex4}
\usepackage{graphics}
\begin{document}
\def\zt{Zlatko Te\v sanovi\' c}
\title{\bf Quasiparticle Thermal Conductivities in a Type-II Superconductor at High Magnetic
Field}
\newcommand{\q}{\bf q} 
\newcommand{\be}{\begin{equation}}
\newcommand{\ee}{\end{equation}}
\author{
 Sa\v {s}a Dukan } 
 \email{dukan@monk.goucher.edu}
 \author{T. Paul Powell} 
 \affiliation{Department of Physics and Astronomy, Goucher College, Baltimore, MD 21204}
 \author{Zlatko Te\v sanovi\' c }
 \affiliation{Department of Physics and Astronomy, Johns Hopkins University, Baltimore, MD 21218} 

\begin{abstract}
~ We present a calculation of the quasiparticle contribution to the longitudinal thermal conductivities $\kappa_{xx}(H,T)$ (perpendicular to the
external field) and $\kappa_{zz}(H,T)$ (parallel to the external field) as well as the transverse (Hall) thermal
conductivity $\kappa_{xy}(H,T)$ of an extreme type-II superconductor in a high
magnetic field  ($H_{c1}\ll H<H_{c2}$) and at low temperatures. In the limit of frequency and temperature
approaching zero ($\Omega \rightarrow 0$, $T\rightarrow 0$),
both longitudinal and transverse conductivities upon entering the superconducting state undergo a reduction
from their respective
normal state values by the factor $(\Gamma /\Delta )^2$, which measures the size of the region at the Fermi surface
containing gapless 
quasiparticle excitations. We use our theory to numerically compute the longitudinal transport coefficient in borocarbide
and A-15 superconductors. The agreement with recent experimental data on LuNi$_2$B$_2$C is very good.
\end{abstract}
\maketitle

\section{INTRODUCTION}

The {\em low-temperature}, {\em high-field} region in the H-T phase diagram of an extreme type-II superconductor is 
the regime where the Landau level quantization of the electronic energies {\it within} the superconducting state 
is well defined, {\it i.e.} the cyclotron energy $\hbar \omega _c >\Delta$, $T$, $\Gamma $ where $\Delta \equiv \Delta (T,H)$ is the 
BCS gap, $T$ is the temperature and $\Gamma \equiv \Gamma (\omega )$ is the scattering rate due to disorder. This regime
should be contrasted with the more familiar opposite limit of 
low magnetic fields and high temperatures where electrons occupy a huge number of Landau levels and where 
the temperature and/or impurity scattering broaden these levels and reduce the significance of Landau quantization. 
With the Landau level structure fully accounted for one discovers a qualitatively new nature of quasiparticle excitations
at high fields: for fields $H$ below but near $H_{c2}$ the quasiparticle spectrum is {\it gapless} at a 
discrete set of points on the Fermi surface. These gapless excitations reflect a coherent quasiparticle 
propagation over many unit cells of a closely packed vortex lattice with fully overlapping vortex cores 
\cite{sasa1,akera,pedro}. The presence of such low-lying excitations makes an s-wave, ``conventional" superconductor in a high magnetic field 
somewhat similar to an anisotropic, ``unconventional" superconductor with 
nodes in the gap. In the low-temperature and high-field regime however, the nodes in the gap reflect the 
{\em center-off-mass} motion of
the Cooper pairs in the magnetic field, in contrast to  d-wave superconducting cuprates where such nodes are due to 
the {\em relative orbital} motion. This gapless behavior in bulk systems is found to persist to  
surprisingly low magnetic fields $H^*\sim 0.2-0.5H_{c2}$. Below $H^*$ the gaps start opening up in the quasiparticle spectrum
and the system eventually reaches the low-field regime of localized states in the cores of isolated, well-separated 
vortices \cite{pedro}. At the present time the strongest evidence for the quantization of quasiparticle orbits
{\em within} the superconducting state comes from observations of the de Haas-van Alphen (dHvA) oscillations in 
various superconducting materials \cite{dhva}. The persistence of the dHvA signal deep within the mixed state, with 
the frequency of 
oscillations still maintaining the normal state value, can be attributed to the presence of a small portion of the Fermi 
surface containing gapless quasiparticle excitations, surrounded by regions where the gap is large \cite{sasa3, maniv}. 

Another useful probe of low-energy excitations in superconductors is measurement of their thermal transport. The simultaneous
measurements of the field dependent longitudinal $\kappa _{xx}(H,T)$ and transverse $\kappa _{xy}(H,T)$ thermal
conductivities are now feasible experimentally and can yield information on both quasiparticle dynamics and the pairing 
mechanism. The dependence of transport coefficients on magnetic field is currently a hotly debated issue in the scientific 
community in light of the experimental observation of field independent plateaus in the longitudinal thermal conductivity 
of high temperature superconductors (HTS) at low fields ($H_{c1}<H\ll H_{c2}$) \cite{plateaus}. The field-independent 
$\kappa _{xx}$ is attributed to the $d_{x^2-y^2}$ pairing mechanism at low fields and to the nodal structure of the 
resulting quasiparticle excitations \cite{marcel,durst}. The presence of propagating gapless quasiparticles in the 
superconducting state at low temperatures and high magnetic fields should also lead to transport properties qualitatively 
different from those found in s-wave superconductors at low fields, where the number of thermally activated quasiparticles 
is exponentially small and the only contribution to the thermal conduction is found along the field direction and 
originates from the bound states within vortex cores. Recently, the thermal conductivity of the borocarbide superconductor
 LuNi$_2$B$_2$C was measured down to $T=70$ mK 
by Boaknin {\it et al.} \cite{boaknin} in a magnetic field perpendicular to the heat current from $H=0$ to above $H_{c2}=7$ Tesla.  In the limit
of $T\rightarrow 0$, a considerable thermal transport was observed in the mixed state  of the superconductor 
($H_{c1}<H\leq H_{c2}$), indicating the presence of delocalized low-energy excitations at the Fermi surface. On the other
hand, no thermal transport was observed at zero field, a result consistent with the s-wave superconducting gap without
 nodes at the Fermi surface.

The purpose of this work is to examine the contribution of low-energy quasiparticles to the thermal transport of 
conventional,
{\it i.e.} extreme type-II superconductors in the regime of high magnetic fields and low temperatures.
 The paper is organized as follows: In Sec. II we develop the Kubo formalism for the transport coefficients within 
 the Landau level pairing mechanism while in Sec. III we incorporate disorder in the Green's function description of a 
three-dimensional superconductor in a high magnetic field.  We use the formalism of Sec. II and III in Sec. IV to
examine both 
longitudinal $\kappa_{xx}(H,T)$ and $\kappa _{zz}(H,T)$ as well as transverse $\kappa _{xy}(H,T)$ conductivities in the 
regime of the freqency $\Omega \rightarrow 0$ and the temperature $T\rightarrow 0$. Finally, in Sec. V we report
on numerical calculations of the thermal transport in the borocarbide LuNi$_2$B$_2$C and the A-15 superconductor V$_3$Si and compare
our theoretical plots with the available experimental data.

\section{KUBO FORMALISM IN THE LANDAU LEVEL PAIRING SCHEME}

Thermal conductivities can be calculated within the framework of the 
Kubo formalism as a linear response of a system to an external perturbation \cite{mahan}
\be
\frac{\kappa_{ij} (\Omega,T)}{T}=-\frac{1}{T^2}\frac{\Im m {\Pi_{ij}^{ret}(\Omega)}}
{\Omega} 
\label{definition}
\ee
where $\Pi_{ij}^{ret}(\Omega)=\Pi_{ij}(i \Omega \rightarrow \Omega + i \delta)$
and 
\begin{eqnarray}
\Pi_{ij}(i \Omega)=\int d^3x_1d^3x_2 \Pi _{ij}(1,2;i\Omega ),
\nonumber\\
\Pi _{ij}(1,2;i\Omega )=-\int_0^{\beta}d\tau e^{i\Omega \tau}<T_{\tau} j_i
(x_1,\tau)j_j(x_2,0)>
\label{tensor}
 \end{eqnarray}
  is the spatially averaged, finite temperature thermal current-current correlation function tensor and $\beta \equiv 1/k_BT$.
 In order to derive the heat current operators $\bf {j}(1)$ and $\bf{j}(2)$ at the space-time point 
 $1=({\bf{x}_1},\tau)$ and $2=({\bf{x}_2},0)$  we follow the standard s-wave derivation in  zero 
 field \cite{ambe} and generalize it to our case of a non-uniform gap at high fields. A similar 
  approach was recently utilized in Ref. 9 for a d-wave superconductor in zero field.
  The heat current carried by the quasiparticles can be computed within the standard variational 
  procedure as 
\be
\bf {j} =\frac{\partial {\cal L}}{\partial (\nabla \psi)}\dot{\psi}+\dot{\psi}^{\dagger}
\frac{\partial {\cal L}}{\partial (\nabla \psi ^\dagger)}
\label{procedure}
\ee
 from the Lagrangian density
\begin{eqnarray}
 {\cal L}=-\frac{1}{2m}\nabla \psi _{\alpha}^{\dagger}\cdot \nabla \psi _{\alpha}
 +\frac{e}{2mci}(\psi _{\alpha}^{\dagger}\nabla \psi _{\alpha}-\nabla \psi _{\alpha}^{\dagger}
 \psi _{\alpha})\cdot {\bf A}
 \nonumber\\
 +\frac{e^2}{2mc^2}|{\bf A}|^2\psi _{\alpha}^{\dagger}\psi _{\alpha}
 -\frac{1}{2i}(
 \psi _{\alpha}^{\dagger}\dot{\psi _{\alpha}}-\dot{\psi _{\alpha}^{\dagger}}\psi _{\alpha})
 \nonumber \\
 -\frac{1}{2}g\psi _{\alpha}^{\dagger}\psi _{-\alpha}^{\dagger}\psi _{-\alpha}\psi _{\alpha}
 \label{lang}
 \end{eqnarray}
 where all the energies are measured with respect to the chemical potential and $\dot{\psi}\equiv\partial\psi/\partial t$. The
 Nambu's two-component field operators $\psi _{\alpha}\equiv \psi _{\alpha}({\bf r})$
 are written in a compact notation for the sake of brevity. In extreme type-II superconductors, as soon as the magnetic field
 satisfies $H\gg H_{c1}(T)$ the vector potential
  ${\bf A}\equiv {\bf A}({\bf r})$ can be safely assumed to be entirely due
  to the external field ${\bf H}= \nabla \times {\bf A}$. This holds over most of the $H-T$ phase diagram. 
 We have used a simple BCS-model point interaction 
 $V({\bf r_1}-{\bf r_2})=-g\delta({\bf r_1}-{\bf r_2})$ in expression (\ref{lang}).  To 
 lowest order in the concentration of impurities, the electron-impurity interaction can be omitted 
 in computing the heat current. The effect of disorder will be included later in the Green's 
 functions for a superconductor. The variational procedure (\ref{procedure}) yields
 \begin{eqnarray}
 {\bf j(1)}=-\frac{1}{2m}\left[\frac{\partial}{\partial \tau_1}(\nabla ' -\frac{e{\bf A'}}{ci})\right.
  \nonumber\\
   \left.   +\frac{\partial}{\partial \tau_1'}(\nabla +\frac{e{\bf A}}{ci})\right]
 \psi _{\alpha}^{\dagger}
 (1')\psi _{\alpha}(1)\vert _{1'=1}
 \label{current}
 \end{eqnarray}
 for the heat current operator. With this 
 definition it is straightforward to calculate the correlation tensor $\Pi _{ij}(1,2;i\Omega)$ within the
 usual Hartree-Fock approximation ({\it i.e.}, bare bubble approximation) defined by 
 \be 
 <T_{\tau} \psi_i(1)\psi_k(2)\psi_{l}^{\dagger}(2')\psi_j^{\dagger}(1')>\rightarrow 
 G_{il}(1,2')G_{kj}(2,1')
 \label{hartree}
 \ee
 where $G_{kl}$ is the Nambu's matrix Green's function. Inserting (\ref{current}) into
 (\ref{tensor}) and with the help of (\ref{hartree}) the current-current correlator
 becomes
 \begin{widetext}
 \begin{eqnarray} 
  \Pi _{ij}(1,2;\Omega )=\frac{1}{4m^2\beta } \sum _{\mu }
  \left[ -i(\Omega +\mu )(
\nabla_{1'} -\frac{e{\bf A(1')}}{ci})
 +i\mu (\nabla_1 +\frac{e{\bf A(1)}}{ci})\right]\times
 \nonumber\\
 \left[ i(\Omega +\mu)(\nabla_2
  +\frac{e{\bf A(2)}}{ci}) 
  -i\mu (\nabla_{2'} -\frac{e{\bf A(2')}}{ci})\right] Tr\left[\tau_3G(1,2',\Omega+
  \mu)\tau_3G(2,1',\mu)\right]_{1\rightarrow 1',2\rightarrow 2'}
 \label{ftrans}
 \end{eqnarray}
 \end{widetext}
 where $\tau_3$ is a Pauli matrix, $\mu=2\pi m/\beta$ are bosonic Matsubara 
 frequencies and $1\equiv {\bf x_1}$.

It was shown in Ref. 1 that the mean-field Hamiltonian  corresponding to the Lagrangian density (\ref{lang})
 can be diagonalized in terms of the basis functions of the Magnetic Sublattice Representation (MSR), 
characterized by the quasi-momentum ${\q}$ perpendicular to the
direction of the magnetic field. 
 The eigenfunctions of this representation 
in the Landau gauge ${\bf A}=H(-y,0,0)$ and belonging to the $m$-th Landau 
level, are
\begin{widetext}
\begin{eqnarray}
\phi _{k_{z},{\q} ,m}({\bf r})=\frac{1}{\sqrt{2^{n}n!\sqrt{\pi }l}} \sqrt{
\frac{b_y}{L_xL_yL_z}}\exp {(ik_z\zeta )}\sum_{k}\exp {(i\frac{\pi b_x}{2a}k^2-ikq_y
b_y)} \times 
\nonumber\\ 
\exp {[i(q_x+\frac{\pi k}{a})x-1/2(y/l+q_xl+\frac{\pi k}{a}l)^2]}H_{m}(\frac
{y}{l}+(q_x+\frac{\pi k}{a})l),
\label{phi}
\end{eqnarray}
\end{widetext}
where $\zeta $ is the spatial coordinate and $k_{z}$ is the momentum along the 
field direction, ${\bf a}=(a,0)$ and ${\bf b}=(b_x,b_y)$ are the unit vectors 
of the triangular vortex lattice, $l=\sqrt{\hbar c/eH}$ is the magnetic
length and $L_xL_yL_z$ is the volume of the 
system. $H_{m}(x)$ is the Hermite polynomial of order $m$. 
Quasimomenta ${\q}$ are restricted to the first Magnetic Brillouin Zone
(MBZ) spanned by vectors ${\bf Q_1}=(b_y/l^2,-b_x/l^2)$ and ${\bf Q_2}=
(0,2a/l^2)$. 

Normal and anomalous Green's functions for a clean superconductor in this representation can be constructed as 
\begin{eqnarray}
G_{11}(1,2;\omega )
=\sum _{n,k_{z},{\q}}
\phi _{n,k_{z},{\q}}(1)\phi _{n,k_{z},{\q}}^{*}(2)
G_n(k_{z},{\q};\omega )
\nonumber \\ 
G_{21}(1,2;\omega)
=\sum _{n,k_{z},{\q}}
\phi ^{*}_{n,-k_{z},-{\q}}(1)\phi ^{*}_{n,k_{z},{\q}}(2)
F_n^*(k_{z},{\q};\omega )
\label{green} 
\end{eqnarray}
where $\omega =(2m+1)\pi/\beta $ are the electron Matsubara frequencies. Similar expressions can be written
for the remaining two Nambu matrix elements.
In writing (\ref{green}) we have taken into account only diagonal (in Landau level index $n$) contributions to the 
Green's functions. This is a good approximation in high magnetic fields 
where $\Delta /\hbar \omega _{c} \ll 1$ and the number of occupied Landau levels 
$n_{c}$ is not too large, which is the case for the extreme type-II systems under consideration. In this situation we are
justified in using the diagonal approximation  
\cite {sasa1,pedro}, in which the BCS pairs are formed by electrons belonging to 
mutually degenerate Landau levels located at the Fermi surface while the 
contribution from Landau levels separated by $\hbar \omega _c$ or
more is included in the renormalization of the effective coupling constant ($g
\rightarrow \tilde{g}(H,T))$ \cite{vafek}. As long as the magnetic field is larger than some critical field $H^*(T)$   
the off-diagonal pairing  does not 
change the qualitative behavior of the superconductor in a magnetic field. The critical field $H^*$ at $T\sim 0$ can
be estimated from the dHvA experiments to be $\sim 0.5H_{c2}$ for A-15 and $\sim 0.2H_{c2}$ for borocarbide
superconductors \cite{dhva}.

When Nambu matrix (\ref{green}) is inserted in Eq. \ref{ftrans} and the space average in (\ref{tensor}) is performed,
the longitudinal $\Pi _{xx}(i\Omega)=\Pi _{yy}(i\Omega)$ and transverse (Hall) 
 $\Pi _{xy}(i\Omega)=-\Pi _{yx}(i\Omega)$
current-current correlation functions become
\begin{eqnarray}
\Pi _{ij}(i\Omega)=\frac{1}{4m^2l^2\beta}\sum_{\omega} \sum_{n,k_z,{\bf q}} (\Omega +2\omega )^2 \frac{n+1}{2}\times
\nonumber \\
Tr[\tau_3G_n(k_z,{\q},i\Omega+
  i\omega)\tau_3G_{n+1}(k_z,{\q},i\omega)
\nonumber \\
  \pm \tau_3G_{n+1}(k_z,{\q},i\Omega+
  i\omega)\tau_3G_{n}(k_z,{\q},i\omega)]
  \label{compact}
  \end{eqnarray}
 where the $+$ sign corresponds to $\Pi _{xx}(i\Omega)$, the $-$ sign corresponds to $i\Pi _{xy}(i\Omega)$ and 
 $\omega = (2m+1)\pi/\beta $ are electronic Matsubara frequencies.
 On the other hand the longitudinal (parallel to the external magnetic field) current-current
 correlation function $\Pi_{zz}(i\Omega )$ becomes
 \begin{eqnarray}
 \Pi _{zz}(i\Omega)=\frac{1}{4m^2\beta}\sum_{\omega} \sum_{n,k_z,{\bf q}} k_z^2 (\Omega +2\omega)^2
 \nonumber \\
 \times Tr\left[\tau_3G_n(k_z,{\q},i\Omega+i\omega)\tau_3G_{n}(k_z,{\q},i\omega)\right].
 \label{compact1}
 \end{eqnarray}
 In order to perform the summation over the Matsubara frequencies $\omega$,
 we introduce a spectral representation for the Nambu matrix $G_{n}(k_z,{\q},\omega)$ as
 \be
 G_{n}(k_z,{\q},\omega)=\int_{-\infty}^{\infty}d\omega_1\frac{A_n(k_z,{\q},\omega_1)}{i\omega -\omega_1}
 \label{specrep}
 \ee
 where the spectral function matrix $A_n(k_z,{\q},\omega)$ is defined as 
 \be
 A_n(k_z,{\bf q},\omega)=-\frac{1}{\pi}Im G_n^{ret}(k_z,{\q},\omega).
 \label{spectral}
 \ee
 When the spectral representation of the Green's functions (\ref{specrep}) is used back in (\ref{compact}) and (\ref{compact1})
 respectively, we obtain 
 \begin{eqnarray}
 \Pi _{ij}(i\Omega)=\frac{1}{4m^2l^2\beta}\sum_{n,k_z,{\bf q}}\int d\omega_1\int d\omega_2\frac{n+1}{2}
 \nonumber \\
 \times Tr[\tau_3
 A_n(k_z,{\q},\omega_1)\tau_3A_{n+1}(k_z,{\q},\omega_2) 
 \nonumber \\
  \pm \tau_3A_{n+1}(k_z,{\q},\omega_1)\tau_3A_{n}(k_z,{\q},\omega_2)]\times S
  \label{withs}
  \end{eqnarray}
  and 
  \begin{eqnarray} 
  \Pi_{zz}(i\Omega )=\frac{1}{4m^2\beta}\sum_{n,k_z,{\bf q}}k_z^2\int d\omega_1\int d\omega_2
  \nonumber \\
  \times Tr\left[\tau_3
 A_n(k_z,{\q},\omega_1)\tau_3A_{n}(k_z,{\q},\omega_2)\right]\times S
 \label{withs1}
 \end{eqnarray}
  where $S$ contains Matsubara sums {\it i.e.}
  \be
  S=\frac{1}{\beta}\sum_{\omega}(\Omega +2\omega)^2\frac{1}{(i\Omega+i\omega-\omega_1)(i\omega -\omega_2)}.
  \label{s}
  \ee
 The sum can be evaluated in the standard way by picking up the contributions from each of the poles of the 
 summand \cite{mahan}. After the analytic continuation
  $i\Omega \rightarrow \Omega +i\delta$ we obtain the retarded function $S_{ret}$
 \be
 S_{ret}=\frac{(2\omega_2+\Omega )^2n_F(\omega _2)-(2\omega_1-\Omega)^2n_F(\omega_1)}{\omega_2-\omega_1+\Omega+i\delta}
 \label{retarded}
 \ee
 where $n_F(\omega)$ is the Fermi function. 
 
 In order to obtain the imaginary part of $\Pi_{ij}(\Omega)$ we need to find an imaginary part of $S_{ret}$ when calculating
 the longitudinal conductivity $\kappa _{xx}(\Omega, T)$ and $\kappa_{zz}(\Omega, T)$. On the other hand, since $S_{ret}$ enters the expression for 
 $i\Pi_{xy}(\Omega)$ in (\ref{withs}) we need to find the real part of $-S_{ret}$. Using the identity
 \be
 \frac{1}{x+i\delta}=P\frac{1}{x}-i\pi\delta(x)
 \label{identity}
 \ee
 and taking the imaginary part of (\ref{s}), we find that the diagonal conductivities become
 \begin{widetext}
 \begin{eqnarray}
 \frac{\kappa_{xx}}{T}=\frac{\kappa_{yy}}{T}=\frac{\pi}{4m^2l^2}\sum_{n}\sum_{k_z,{\q}}\int_{-\infty}^{+\infty}d\omega \frac
 {(2\omega +\Omega)^2}{T^2}\frac{n_F(\omega)-n_F(\omega+\Omega)}{\Omega}\times
 \nonumber\\
 \frac{n+1}{2}Tr\left[\tau_3
 A_n(k_z,{\q},\omega+\Omega)\tau_3A_{n+1}(k_z,{\q},\omega)+\tau_3A_{n+1}(k_z,\q,\omega+
  \Omega)\tau_3A_{n}(k_z,{\q},\omega)\right]
  \label{diagonal}
  \end{eqnarray}
  \end{widetext}
 and 
 \begin{widetext}
  \be
  \frac{\kappa_{zz}}{T}=\frac{\pi}{4m^2}\sum_{n,k_z,{\bf q}}k_z^2\int_{-\infty}^{+\infty} d\omega
  \frac{(2\omega +\Omega)^2}{T^2}\frac{n_F(\omega)-n_F(\omega+\Omega)}{\Omega}
  Tr\left[\tau_3
 A_n(k_z,{\q},\omega +\Omega)\tau_3A_{n}(k_z,{\q},\omega)\right]
 \label{diagonalz}
 \ee
 \end{widetext}
 Similarly, taking the real part of (\ref{s}) with the help of (\ref{identity})
  yields for the off-diagonal conductivity:
 \begin{widetext}
  \begin{eqnarray}
  \frac{\kappa_{xy}}{T}=-\frac{\kappa_{yx}}{T}=\frac{1}{4m^2l^2}
  \sum_{n}\sum_{k_z,\q}\int_{-\infty}^{+\infty}d\omega \frac{(2\omega +\Omega)^2}{T^2}
  \frac{n_F(\omega)-n_F(\omega+\Omega)}{\Omega}\times
  \nonumber\\
  \frac{n+1}{2}Tr\left[\tau_3B_{n+1}(k_z,{\q},\omega+\Omega)\tau_3A_{n}(k_z,{\q},\omega)
  -\tau_3B_n(k_z,{\q},\omega+\Omega)\tau_3A_{n+1}(k_z,{\q},\omega)\right]
  \label{offdiagonal}
  \end{eqnarray}
  \end{widetext}
  where the function $B_{n}(k_z,{\q},\omega+\Omega )$ is defined as 
  \be
  B_{n}(k_z,{\q},\omega+\Omega )=\int_{-\infty}^{+\infty}d\omega_1\frac{A_{n}(k_z,{\q},\omega_1)}{\omega -\omega_1+\Omega}.
  \label{bfunction}
  \ee
 
 \section{GREEN'S FUNCTIONS IN THE PRESENCE OF DISORDER}
  
 Before further discussing expressions (\ref{diagonal}), (\ref{diagonalz}) and (\ref{offdiagonal}) we should go back to the question 
 of spectral functions or,
 alternatively, Green's functions for the superconductor in a magnetic field. The Green's function for the clean superconductor
 can be easily found following Ref. 14 with their ``Fourier transforms" in the quasi-momentum space expressed 
 in the Nambu formalism as
 \be
 {\cal G}_n=\frac{1}{(i\omega)^2-E_{n}(k_z,{\q})}\left(\begin{array}{cc}
 i\omega+\epsilon_{n}(k_z) &-\Delta_{nn}({\q})\\
 -\Delta_{nn}^*({\q})& i\omega-\epsilon_{n}(k_z)
 \end{array}\right)
 \ee
 where
\begin{eqnarray}
E_{n,p}(k_{z},{\q})=p\hbar\omega_c\pm \sqrt{\epsilon _{n}^{2}(k_z)+|\Delta _{n+p,n-p}({\q})|^{2}}
\nonumber\\
\epsilon _{n}(k_z)=\frac{\hbar ^2k_{z}^{2}}{2m}+\hbar \omega _{c}(n+1/2)-\mu
\label{epsilon}
\end{eqnarray}
is the quasiparticle excitation spectrum of the superconductor in high magnetic field near 
points $k_z=\pm k_{Fn}=\sqrt{2m(\mu-\hbar\omega_c(n+1/2))/\hbar^2}$ calculated  within the diagonal
 approximation \cite{sasa1,pedro}, where $\Delta/\hbar\omega_c\ll1$. For the quasiparticles near the Fermi surface 
 ($k_z\sim
 k_{Fn}$)
  it suffices to consider only the $E_{n,p=0}$ bands.   The gap, $\Delta _{nn}({\q})$, which in the MSR representation can 
  be written as
\begin{eqnarray}
\Delta _{nm}({\q} )=\frac{\Delta }{\sqrt{2}}\frac{
(-1)^{m}}{2^{n+m}\sqrt{n!m!}}\sum_kexp(i\pi \frac{b_x}{a}k^2)\times
\nonumber \\
exp(2ikq_yb_y-(q_x+\frac{\pi k}{a})^2l^2)
H_{n+m}[\sqrt{2}(q_x+\frac{\pi k}{a})l],
\label{delta}
\end{eqnarray} 
turns to zero on the Fermi surface at the set of points in the MBZ with a strong linear dispersion in $q$. The excitations
from other, $p\neq 0$ in (\ref{epsilon}), bands are gapped by at least a cyclotron energy and their contribution to the 
quasiparticle transport can be neglected at low temperatures ($T\ll\Delta(T,H)\ll\hbar\omega_c$). Once the off-diagonal
 pairing in (\ref{green}) is included, the excitation spectrum cannot be written in the  
simple form (\ref{epsilon}) and a closed analytic expression for the superconducting Green's function cannot be found.
Nevertheless, when these off-diagonal terms are treated pertubatively as in Ref. 3 the qualitative behavior of 
the quasiparticle excitations, characterized by
the nodes in the MBZ, remains the same. This statement is correct in all orders of the perturbation theory and therefore
is exact as long as the pertubative expansion itself is well defined {\it i.e.} as long as $H>H^*(T)$.   
Once the magnetic field is lowered below $H^*$, gaps start
opening up at the Fermi surface signalling the crossover to the low-field
regime of quasiparticle states localized in the cores of widely separated vortices 
\cite{norman1}.

In a dirty but homogenous superconductor with the coherence length $\xi$ much longer than the effective distance 
$\xi_{imp}$ over which the impurity potential changes ($\xi/\xi_{imp}>>1$), the superconducting order parameter is
not affected by the impurities and still forms a perfect vortex lattice. 
For such a system, the bare Green's 
function in (\ref{green}) is dressed via scattering through the diagonal
(normal) self-energy $\Sigma ^N(i\omega)$ and off-diagonal (anomalous) self-energy $\Sigma^{A}_{nn}({\q},i\omega)$ 
.\cite{sasa2} A dressed Green's function is obtained by replacing $\omega$ with $\tilde{\omega}$ and 
$\Delta_{nn}({\q})$ with $\tilde{\Delta}_{nn}({\q})$ in (\ref{green}) where
\begin{eqnarray}
i\tilde{\omega}\equiv i\omega - \Sigma ^N(i\omega)
\nonumber\\
\tilde{\Delta}_{nn}({\q})\equiv\Delta_{nn}({\q})+ \Sigma^{A}_{nn}({\q},i\omega).
\label{dressed}
\end{eqnarray}
In order to calculate the spectral functions in (\ref{diagonal}) the analytical continuation 
should be performed so that $G_{ret}(k_z,{\q},\omega)=G(k_z,{\q},i\omega\rightarrow\omega
+i\delta)$ where $\Sigma_{ret}^{N,A}(\omega)=\Sigma^{N,A}(i\omega\rightarrow\omega
+i\delta)$ with the impurity scattering rate in the superconducting state 
defined as $\Gamma(\omega)=-Im\Sigma_{ret}^{N}
(\omega)$. It was shown by the authors in Ref. 14 that the  anomalous self energy
does not qualitatively change the form of the gap function $\Delta_{nn}({\q})$ at low 
energies and therefore $\Sigma^{A}_{nn}({\q},\omega)$ will be neglected in further calculations.
 At the same time, the real part of
the normal self energy $\Sigma^{N}(\omega)$ can be either neglected or absorbed into 
$\epsilon _n(k_z)$. 

\section{THERMAL CONDUCTIVITIES IN  $T\rightarrow 0$ LIMIT}
 
  We are interested in calculating thermal conductivities in (\ref{diagonal}) and (\ref{offdiagonal}) 
in the limit of $\Omega\rightarrow 0$ and small $T$ such that $T\ll \Gamma (\omega)$.
In the limit of $\Omega\rightarrow 0$ the difference of Fermi functions in (\ref{diagonal})
 becomes
\be
\frac{n_F(\omega +\Omega)-n_F(\omega)}{\Omega}\rightarrow \frac{\partial n_F}{\partial \omega}.
\label{limit}
\ee
This function  is sharply peaked around $\omega =0$ at very low temperatures so that we are
justified in expanding the integrand in (\ref{diagonal}) and (\ref{offdiagonal}) around $\omega =0$
 up to second order in $\omega $ and setting 
the scattering rate to a constant $\Gamma =\Gamma (\omega =0)$. In the high-field superconductors
the largest contribution to the thermal conductivity comes from the quasiparticle excitations
at the Fermi surface with momenta ${\bf q}$ such that $\Delta({\q}) \leq max(T,\Gamma )$ 
while the excitations gapped by large $\Delta ({\q})$ give exponentially small contributions.
Therefore, in order to simplify the integration over the MBZ and summation over the Landau level 
index in (\ref{diagonal}) and (\ref{offdiagonal}) we linearize the excitation spectrum (\ref{epsilon}) around nodes 
at the Fermi surface \cite{sasa3}. This enables us to obtain approximate but analytic expresions for thermal 
transport coefficients which capture the qualitative behavior near $H_{c2}$. Keeping this in mind and with
the help of the Sommerfeld expansion \cite{achroft} the longitudinal 
conductivity $\kappa_{xx}(H,T)=\kappa_{yy}(H,T)$ and $\kappa_{zz}(H,T)$ at low temperatures  become
\be
\frac{\kappa _{xx}(H,T)}{\kappa_{xx}^N(H,T)}=(\frac{4}{\pi}-1)
\left(\frac{\Gamma }{\Delta}\right)^2 +
\frac{7\pi^2}{5}(1-\frac{3}{\pi})\left(\frac{k_BT}{\Delta}\right)^2
\label{fdia1}
\ee
and
\be
\frac{\kappa _{zz}(H,T)}{\kappa_{xx}^N(H=0,T)}=
\left(\frac{\Gamma }{\Delta}\right)^2 +
\frac{7\pi^2}{15}\left(\frac{k_BT}{\Delta}\right)^2.
\label{fdiazz}
\ee
 $\kappa_{xx}^N(H,T)$ is the thermal conductivity of the normal metal in a
 magnetic field \cite{abrikosov}
\be
\kappa _{xx}^N(H,T)=\frac{\pi ^2}{3}\frac{n_e}{2m^*\Gamma }\frac{4\Gamma ^2 }{(\hbar \omega _c)^2+4\Gamma ^2}T
\label{normald}
\ee
where $n_e=\frac{1}{2\pi l^2}\sum_n k_{Fn}$ is electronic density in the system.  
On the other hand the transverse conductivity $\kappa_{xy}(H,T)=-\kappa_{yx}(H,T)$ becomes  
 \be
 \frac{\kappa _{xy}(H,T)}{\kappa _{xy}^N(H,T)}=(\frac{4}{\pi}-1)
\left(\frac{\Gamma }{\Delta}\right)^2 +
\frac{7\pi^2}{5}(1-\frac{3}{\pi})\left(\frac{k_BT}{\Delta}\right)^2
 \label{foff1}
 \ee
 where $\kappa_{xy}^N(H,T)$ is the off-diagonal thermal conductivity of the normal metal in
 magnetic field \cite{abrikosov}
 \be
 \kappa_{xy}^N(H,T)=\frac{\pi ^2}{3}\frac{n_e}{m^*\omega _c}\frac{(\hbar \omega _c)^2}{(\hbar \omega_c)^2+4\Gamma ^2}T.
 \label{normaloff}
 \ee
 
 Relations (\ref{fdia1}), (\ref{fdiazz}) and (\ref{foff1}) obtained when $\Gamma/\Delta \ll 1$ within 
 the ``linearized spectrum aproximation" tell us 
 that there is still a considerable thermal transport in the mixed state of the superconductor. This is in stark contrast
  to the exponential suppresion of transport 
 characteristic of an s-wave 
 superconductor in zero field. Furthermore, relations (\ref{fdia1}), (\ref{fdiazz}) and (\ref{foff1}) indicate that 
 when passing
 from the normal to the superconducting state, both longitudinal and transverse transport coefficients $\kappa /T$ 
 are reduced from their 
 respective normal state values by the factor $\sim (\Gamma /\Delta )^2$ (the term linear in $(T/\Delta)^2$ is negligible 
 at low temparatures even for very clean superconductors). The factor $\sim (\Gamma /\Delta )^2$ measures the fraction
 of the Fermi surface $\cal G$ 
 containing gapless quasiparticle excitations at $T=0$.  
 The size of $\cal G$ is determined by both the total number
 of nodes in the excitation spectrum (\ref{epsilon}) and the areas in different branches where the BCS gap $\Delta$ is 
 very small but not 
 necessarily zero. This result, obtained here for the thermal coefficients, is consistent with the behavior of some
  other superconducting 
 observables that measure the presence of low-energy excitations at the Fermi surface. One such experimentally
  confirmed behavior is 
 the reduction of 
 the de Haas-van Alphen (dHvA) oscillation's  amplitude in both A-15 and borocarbide superconductors when the 
 sample becomes superconducting \cite{dhva}. 
 The drop in
 the overall amplitude in passing from the normal to the superconducting state reflects the presence of a small portion of the 
 Fermi surface $\sim \cal G$ containing coherent gapless excitations while the rest is gapped by large $\Delta$. \cite{sasa3}
   
\section{COMPARISON WITH EXPERIMENT}

Recently, the longitudinal thermal conductivity of the borocarbide superconductor LuNi$_2$B$_2$C was measured down to $T=70$ mK 
by Boaknin {\it et al.} \cite{boaknin} in a magnetic field from $H=0$ to above $H_{c2}=7$ Tesla.  In the limit
of $T\rightarrow 0$, a considerable thermal transport is observed in the mixed state  of the superconductor 
($H_{c1}<H\leq H_{c2}$), indicating the presence of delocalized low-energy excitations at the Fermi surface. The authors argue that
this observation is strong evidence for a highly anisotropic gap function in LuNi$_2$B$_2$C, possibly with nodes. On the other 
hand, no sizable thermal conductivity was observed in zero field, the result expected for a superconducting gap
without nodes.

The Boaknin {\it et al.} result is consistent with
the observation of dHvA oscillations down to fields $H^*\sim H_{c2}/5$ in YNi$_2$B$_2$C, a close cousin of LuNi$_2$B$_2$C
 as well as in V$_3$Si where the oscillations persist down to fields $H^*\sim H_{c2}/2$.\cite{dhva} 
 It was shown by us that the drop in the dHvA amplitude observed in these experiments can be atributed to the 
quantization of quasiparticle
 orbits within the superconducting state which results in the formation of nodes in the gap \cite{sasa3}. This quantum
  regime behavior in fields $H^*<H<H_{c2}$
 is due to the center-off-mass motion of the Cooper pairs, in contrast to the d-wave or anisotropic s-wave where nodes in the
 gap are due to the relative orbital motion. Therefore it makes sense to compare the theory developed in this paper 
 with the 
 experimental data in Ref. 10.  
 %
 A quick check tells us that the expression (\ref{fdia1}), where $\Gamma (H)/\Delta (H)<1$, does not hold through 
 the entire range of fields
 used in the experiment. Therefore, we {\em numerically} compute the longitudinal thermal conductivity directly
 from Eq. 19, without 
 using any additional aproximations, for both the borocarbide 
 superconductor LuNi$_2$B$_2$C as well as for the A-15 superconductor V$_3$Si.  
 
 In the
 limit of $\Omega \rightarrow 0$ and $T\rightarrow 0$ the expression (\ref{diagonal}) yields
 \begin{widetext}
 \begin{equation}
 \frac{\kappa}{T}=\frac{\pi }{12m}\frac{\Gamma ^2}{\hbar \omega_c}\sum_{n}^{n_c}(n+1)\sum_{k_z,{\q}}\frac{1}{E_{n,p=0}^2(k_z,
 {\q})+\Gamma ^2}\cdot \frac{1}{E_{n+1,p=0}^2(k_z,{\q})+\Gamma ^2}
 \label{computer}
 \end{equation}
 \end{widetext}  
 where the number of Landau levels involved in superconducting pairing $n_c=E_F/\hbar \omega_c$ varies as a function of 
 magnetic field. In the borocarbide superconductor LuNi$_2$B$_2$C $n_c$ can be estimated as $n_c\sim 33$ at $H_{c2}$ and
  $n_c\sim 1147$ at field of $H=0.2$ Tesla (these 
 numbers were obtained using an effective mass of $0.35m_e$ and Fermi velocity $v_F=2.76\times 10^7$ cm/s as reported in 
 Ref. 18). On the other hand, the number of occupied Landau levels in the A-15 superconductor V$_3$Si is much larger:
  $n_c\sim 241$   at $H_{c2}=18.5$ Tesla and $n_c\sim 4470$ at $H=1$ Tesla 
  (we used an effective mass of $1.7m_e$ and Fermi velocity $v_F=2.8\times 10^7$ cm/s from Ref. 19 in this estimate).
 The scattering rate $\Gamma=\Gamma (\omega=0)$ in (\ref{computer}) is, in general, modified relative to the normal state
 scattering rate $\Gamma _0$ when the system becomes superconducting. Indeed, the self-consistent calculation of $\Gamma$
 in Ref. 14
 gives $\Gamma (H)=\sqrt{\Gamma _0 \Delta (H) /2}$. We assume $\Delta(H)=\Delta \sqrt{1-H/H_{c2}}$ which is a good 
 approximation for the range of fields used in the experiment. 
 
 The dashed line in Fig. 1 shows the magnetic field dependence of the quasiparticle thermal conductivity $\kappa /T$ for the 
  borocarbide superconductor LuNi$_2$B$_2$C in the limit
 of $T\rightarrow 0$ obtained by the numerical evaluation of (\ref{computer}), where values for the BCS gap $\Delta$ and 
 normal state inverse scattering rate $\Gamma _0$ are taken from Ref. 10. The full circles 
 in Fig. 1 are the experimental data of Boaknin 
 {\it et al.} \cite{boaknin} for the same superconductor. The full line in Fig. 1 shows 
 the theoretical plot obtained by numerical evaluation of (\ref{computer}) for  the A-15  superconductor V$_3$Si 
 where values for 
 $\Delta$ and $\Gamma _0$ are taken from Ref. 19. 
 There is a significant difference in the behavior of $\kappa/T$ in these two superconducting systems characterized by  
 much smaller thermal transport in V$_3$Si  when compared to the transport in LuNi$_2$B$_2$C in the same range of magnetic
 fields. 
 This observation indicates that the number of gapless or near gapless
 excitations at the Fermi surface in the V$_3$Si is very small at $H\ll H_{c2}$. In order to understand this difference,
 one has to notice that in magnetic fields $H<H^*=0.5H_{c2}$ the number of occupied
 Landau levels in V$_3$Si system is huge ($\leq 4500$) and that V$_3$Si is away from the regime of coherent gapless
 excitations of high fields. On the other hand, the number of occupied Landau levels in LuNi$_2$B$_2$C is much smaller 
 ($\leq 1000$) and it seems that there are still many gapless excitations left at low fields. It is suprising though
 that significant thermal transport exists down to $H\sim 0.015H_{c2}$, a field much smaller than the critical field
 $H^*=0.2H_{c2}$ for this system, where most of the quasiparticle spectrum should be gapped. Note, however, 
 that a possible source for such a significant transport at low fields might be the highly anisotropic s-wave gap function 
 in LuNi$_2$B$_2$C, as suggested in Ref. 10. If there is such anisotropy along one or more directions on the Fermi surface,
 the range of validity of our theory may be extended to fields lower than the simple estimate for $H^*$.\cite{comment}
 In this regard, we alert the reader that at the lowest fields in Fig. 1 ($H\sim 0.04H_{c2}$)
 our theory is stretched to its very limits and its quantitative accuracy diminishes. 
 
\begin{figure}
\scalebox{0.5}{\includegraphics{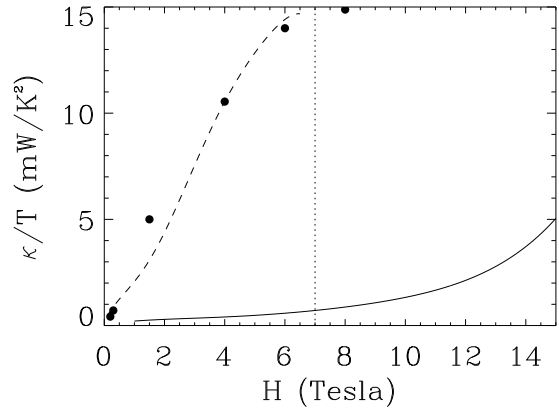}}
\caption{\label{fig1}Magnetic field dependence of the quasiparticle longitudinal thermal conductivity computed from Equation (\ref{computer})
for LuNi$_2$B$_2$C (dashed line) and V$_3$Si (full line). Full circles represent experimental data of Boaknin {\it et al.}
\cite{boaknin} The vertical dotted line indicates the normal-superconducting transition at $H_{c2}=7$ Tesla for LuNi$_2$B$_2$C.
The upper critical field for V$_3$Si at $H_{c2}=18.5$ Tesla is not shown. For LuNi$_2$B$_2$C we have used experimentally determined values 
for $\Delta =4.4$ meV and $\Gamma_0=0.5\Delta $ from Ref. 10 as well as the effective mass 
$m^*=0.35m_e$ from Ref. 18.
For V$_3$Si, $\Delta =2.6$ meV, $\Gamma_0=0.61$ meV and $m^*=1.7m_e$ were taken from Ref. 19.}
\end{figure}

\section{CONCLUSIONS}

In this paper we develop expressions for the longitudinal and transverse quasiparticle thermal conductivities for
an extreme
type-II superconductor in a magnetic field. 
We utilize the Landau level formalism of superconducting pairing in a magnetic field to obtain, within the Kubo 
mechanism of linear response to an external
 perturbation, thermal currents 
perpendicular and parallel to the external magnetic field. From there, current-current correlation functions are introduced
within the Matsubara finite temperature mechanism in order to derive closed expressions for thermal conductivities 
$\kappa _{ij}(\Omega ,T)$. We examine the transport coefficients $\kappa _{ij}/T$ in the limit of $\Omega \rightarrow 0$
and $T \rightarrow 0$ and find that there is considerable thermal transport in the mixed state of a superconductor 
with an s-wave symmetry due to the creation of gapless exitations in the magnetic field. This is in contrast to the zero 
field thermal transport which is exponentially small for an s-wave
superconductor with no nodes in the gap. Furthermore, when passing from the normal to the superconducting state the 
thermal coefficients 
$\kappa _{ij}/T$ become reduced
with respect to their normal state values by a factor $\sim (\Gamma/\Delta)^2$ which measures the fraction of the Fermi
surface that contains coherent gapless or near gapless excitations in a magnetic field. In this respect, thermal conductivities
behave similarly to the dHvA oscillations in which the amplitude is also reduced at the superconducting transition.
Finally, we numerically compute the longitudinal thermal conductivity for two realistic superconducting systems, the 
borocarbide LuNi$_2$B$_2$C and the A-15 superconductor V$_3$Si. The thermal transport in  LuNi$_2$B$_2$C is much larger 
in magnitude
than the thermal transport of V$_3$Si at the same field. This result indicates that the
borocarbide LuNi$_2$B$_2$C might still be in the regime of delocalized quasiparticle states even at fields much 
lower than the critical field $H^*\sim 0.2H_{c2}$ (estimated from the dHvA experiments \cite{dhva,comment}). The agreement of our theoretical
plot with the experimental data for LuNi$_2$B$_2$C taken by Boaknin {\it et al.} \cite{boaknin} over a wide range of fields used in the
experiment is suprisingly good. 

\section*{ACKNOWLEDGEMENT}

The authors would like to thank E. Boaknin and L. Taillefer  for sharing their knowledge of 
the subject with us. This work is supported by a grant from
the Research Corporation (SD and TPP) and by the NSF Grant DMROO-94981 (ZT).


\begin{thebibliography}{99}
\bibitem{comment} In a clean s-wave superconductor $x=H^*/H_{c2}$ can be estimated as a solution of the equation 
$x^3=\frac{2\Delta^4}{\pi E_F(\hbar \omega _{c2})^3}(1-x)^2$ and yields $H^*\sim0.2-0.5H_{c2}$ for the systems in question 
(see Ref. 4). This estimate depends on the value of the s-wave gap function and can be much smaller than $0.2-0.5H_{c2}$
if the value of minimum gap in a strongly anisotropic s-wave case is significantly different from the accepted BCS value.
\end{thebibliography}
\end{document}